\documentclass[prd,preprintnumbers,12pt]{revtex4}
\usepackage{epsfig}
\usepackage{graphicx}
\usepackage{bm}

\newcommand{\be}{\begin{equation}}
\newcommand{\dd}{\displaystyle}
\newcommand{\ee}{\end{equation}}
\newcommand{\bea}{\begin{eqnarray}}
\newcommand{\eea}{\end{eqnarray}}

\begin{document}\preprint{{\bf BARI-TH 514/05}}
\title{Nambu Jona-Lasinio Model of $\bar q q$ Bose Einstein Condensation and pseudogap phase}
\author{P.Castorina$^{(a,b)}$, G. Nardulli$^{(c,d)}$, D. Zappal\`a$^{(a,b)}$}
\affiliation{$^{(a)}$ Department of Physics, University of
Catania, Italy\\ $^{(b)}$ INFN-Catania, Italy  \\
$^{(c)} $Department of Physics and TIRES Center, University of
Bari, Italy \\ $^{(d)} $ INFN-Bari, Italy}

\begin{abstract}We show the existence of a pseudogap phase in the
Nambu Jona-Lasinio model of quark interactions. In the pseudogap
phase chiral symmetry is restored but $q\bar q$ pseudoscalar mesons
still exist and they are massive. Such a behavior is intermediate
between a BCS superconductor and a Bose Einstein Condensate. We
suggest the relevance of this phenomenon for an understanding of
recent lattice QCD and experimental data.

PACS number: 11.30Rd\end{abstract} \maketitle
\section{Introduction \label{sec:0}}

Our current understanding of nuclear phenomena indicates that, at
finite temperature, hadronic matter undergoes a phase transition
to deconfined gluons and quarks. Quantum-Chromo-Dynamics (QCD)
lattice calculations \cite{karsh1} strongly support the conclusion
that, at some critical temperature, a transition to the
quark-gluon plasma (QGP) phase occurs; this temperature is
numerically equal to the critical temperature for the restoration
of the chiral symmetry.

The earliest suggested signature of QGP was the strong suppression
of the light and heavy $q\bar q$ bound states for temperatures
larger than critical temperature \cite{Shuryak:1978ij}. However,
more recent lattice results \cite{karsh2}, using the Maximal
Entropy Method, have found that  mesonic bound states, light or
heavy, actually persist for temperature at least a factor of two
larger than the critical temperature and the quasiparticle mass
turns out to be still  quite large  \cite{karsh3}. Furthermore,
the RHIC data on the radial and elliptic flows \cite{rhic1} can be
explained  by partonic cascades \cite{molnar} and viscosity
corrections \cite{tea} only if the partonic cross section is about
50 times larger than the perturbative QCD calculations, which
indicates a strong coupling regime.

Some models have been proposed to describe this strongly
interacting  phase \cite{shu1,simonov}. The general feature
resulting from both  theory and experiment is the appearance of
two different temperatures: $T_c$ and $T^*$ ($T_c<T^*$). The lower
temperature $T_c$ should be associated with deconfinement and
chiral symmetry restoration, the upper $T^*$  is related to the
decoupling of the bound states from the spectrum \cite{karsh2}.

The presence of two temperature scales is an interesting phenomenon that has an
analogue  in high temperature superconductors. The nature of the phase
transition for these systems is still matter of debate, see, for a recent
review \cite{Loktev:2000ju}. An established experimental fact seems however to
be the existence in these superconductors of a  pseudogap \cite{Emery:1995mn},
which is a depletion of the single particle density of states around the Fermi
level.  Another characteristic feature of the high T$_c$ superconductors, is
 their  coherence length $\xi_0$ which is much smaller than in ordinary
 superconductors. Both features might be the phenomenological manifestations
of a crossover from the Bardeen-Cooper-Schrieffer (BCS) behavior
of the ordinary superconductors to a Bose Einstein Condensate
(BEC) behavior \cite{Griffin}. In this scheme at a certain
critical temperature a pseudogap phase is reached where the
quasiparticles are still gapped although phase coherence and long
range order are lost.  Superconducting fluctuations might be
responsible for this phenomenon, though other interpretations are
also possible \cite{Loktev:2000ju}.  The relation between the
pseudogap transition and field fluctuations has been qualitatively
elucidated  in the  Nambu  Jona-Lasinio (NJL) model
\cite{Kleinert:1999wm,Babaev:2000fj}; moreover it has been
discussed in \cite{Zarembo:2001wr}  in the framework of a
semi-phenomenological constituent quark model,  in \cite{kitazawa}
in the case of color superconductivity and in \cite{schuck} for
low density nuclear matter.

In this paper we analyze in more detail  the dynamics of this NJL
pseudogap phase, described by  fermionic degrees of freedom and/or
by the equivalent bosonic system, including also finite density
effects. The picture that emerges from an explicit numerical
calculation can be summarized as follows:

i) Chiral symmetry is broken in the chiral limit, at  $T <T_c$ and
small density. The scalar meson, $\sigma$, is massive and the
pseudoscalar meson, $\pi$, is the massless Nambu Goldstone Boson
(NGB).

ii) The temperature (and/or the density) restores the chiral
symmetry and the order parameter $ <\bar \psi \psi> $ goes to zero
at $T_c$. The signal of the restoration of the chiral symmetry,
due to the background fluctuations, is given by the equal mass  of
the scalar and pseudoscalar mesons for $T > T_c$. Therefore the
NGB acquires a mass, while remaining in the spectrum. In this
phase the constituent fermion mass remains finite (pseudogap
regime) and there is a strong fermion-meson coupling.

iii) Above  another critical temperature, $T^* > T_c$, the
constituent fermion mass goes to zero and  the mesons decouple
from the physical spectrum.

The paper is organized as follows. In section \ref{sec:1}  we
present our formalism. In section \ref{sec:2} we discuss the
bosonization of the NJL model and in section \ref{sec:3} we
compute the stiffness. In section \ref{sec:4} we compute the
pseudogap and in section \ref{conclusions} we draw our
conclusions.

\section{Constituent quark mass and pion decay
constant \label{sec:1}}

We use\be{\cal L}={\cal L}_0+{\cal L}_4\ee with massless quarks at
 finite chemical potential in two flavors ($N_f=2$). Here \be
{\cal L}_0=\bar\psi(i\partial_\nu\gamma^\nu+\mu\gamma_0)\psi\ee
while ${\cal L}_4$ gives the Nambu Jona-Lasinio (NJL) coupling
(\cite{Nambu:1961fr} and for reviews \cite{Klevansky:1992qe},
\cite{Hatsuda:1994pi}): \be {\cal L}_4=
\frac{G_0}{2N_c}\left[({\bar \psi}\psi)^2+({\bar
\psi}i\gamma_5{\bm\tau}\psi)^2\right]\ .\ee In the sequel we
neglect the so called exchange terms \cite{Klevansky:1992qe}. In
the Mean Field Approximation (MFA) the constituent quark mass is
obtained by the gap equation, which at $T=\mu=0$ is as follows
\be\label{gap0} m_q=4N_fN_c\frac{G_0}{2N_c}\int_\Lambda
\frac{d^3p}{(2\pi)^3}\frac {m_q}{\sqrt{p^2+m^2_q}}\ ,\ee where
$\Lambda$ is a 3D cutoff. One can note the relation between the
constituent quark mass and the chiral condensate  \be m_q=
-\,\frac{G_0N_f}{N_c}\langle 0| \bar u u|0\rangle\ .\label{MFA}\ee

Going beyond the MFA means  considering diagrams of order
$\dd\left(\frac{G_0}{2N_c}\right)^2$, see e.g.
\cite{Dimitrovich:1993}. Clearly the smaller $N_c$, the larger the
role of the fluctuations. As a consequence, an effective momentum
dependent quark mass is introduced $M_{eff}(p)$. The momentum
dependence varies with the model, but for the model we use here,
which corresponds to model $A$ of \cite{Dimitrovich:1993},
$M_{eff}(p)$ is almost constant, with variations of a few percent
in the whole $p$ range. Therefore we will neglect these effects
altogether. Moreover it can be observed that the reliability of
the mean field gap equation to evaluate the constituent quark mass
also in a region where the fluctuations are relevant (as for
example for small $N_c$), has been  explained by Witten
\cite{witten} in exactly solvable models in 1+1 dimension and
numerically checked by lattice calculations in  2+1 dimensional
 models \cite{Hands:2001cs}.

The gap equation at finite $T$ and $\mu$ can be obtained by
considering a sum over Matsubara frequencies, by the substitution
of the energy integration variable $p^0$ with $i\omega_n-\mu$,
where $\mu$ is the quark chemical potential and $\omega_n=\pi
T(2n+1)$.

Eq. (\ref{gap0}) relates the constituent quark mass $m_q$ at
$T=\mu=0$ to the NJL coupling $G_0$ and the cutoff $\Lambda$. Its
generalization for finite $T$ and $\mu$ provides the $T$ and
$\mu$-dependent constituent quark mass in the mean field
approximation: $m_q(T,\mu)$. By getting rid of the NJL coupling
constant, we get $m_q(T,\mu)$ from the constituent mass at
$T=\mu=0$ as follows:\be\label{eq:1.4}
0=\int_\Lambda\frac{d^3p}{(2\pi)^3}\left[\frac{1}{\sqrt{p^2+m^2_q}}
-\frac{\sinh\,y}{\epsilon\left(\cosh\,y+\cosh
\,x\right)}\right]\label{eq:7}\ee where \be x=\frac\mu{T}\
,\hskip1cm
y=\frac{\epsilon}{T}\,,\hskip1cm\epsilon=\frac{\sqrt{p^2+m^2_q(T,\mu)}}{T}\
.\label{eq:9}\ee
 The pion decay constant at $T=\mu=0$ is
\be\label{eq:1.1}
f_\pi^2=-4im_q^2N_c\int_\Lambda\frac{d^4p}{(2\pi)^4}\frac{1}
{(p^2-m_q^2+i\epsilon)^2}\ .\ee One can use this equation to get
$\Lambda$ by fixing
 the constituent quark mass. Using  $m_q=300$ MeV as an
input, from $f_\pi=93.3$ MeV one gets $\Lambda=675$ MeV.

 We
define the  $T^*=T^*(\mu)$ temperature by $ m_q(T^*,\mu)=0$;
numerically one finds, at $\mu=0$, $T^*=185 $ MeV. For generic
values of $\mu$ the result can be obtained by Eq. (\ref{eq:7}).
The generalization of Eq. (\ref{eq:1.1}) to finite $T$ and $\mu$
is straightforward.

\section{Bosonization of the NJL\label{sec:2}}
Let us first work at $T=\mu=0$. As is well known the NJL model can
be made equivalent to  the linear $\sigma$ model
\cite{Eguchi:1976iz} (see also \cite{Klevansky:1992qe}). One
introduces  fields $\sigma$ and $\bm \pi$  by writing the generating
functional as follows \bea Z&=&\int[d\psi][d\bar\psi]\
\exp\left\{\dd\int dx ({\cal L}_0 +{\cal L}_4)\right\}=
\int[d\psi][d\bar\psi][d\sigma][d\pi^i]
\cr&\times&\exp\left\{\dd\int dx\left(
\bar\psi(i\partial_\nu\gamma^\nu+\mu\gamma_0-g_0(\sigma+i\gamma_5{\bm
\tau}\cdot{\bm\pi}))\psi-\frac{g_0^2N_c}{2G_0}\left[\sigma^2+{\bm
\pi}^2\right] \right)\right\} \label{13bis}\eea where \be
g_0=\frac{m_q}{f_\pi}\label{GT}\ee is the meson-quark coupling
constant. Eq. (\ref{GT}) is the analogous of the Goldberger-Treiman
relation. By bosonization and derivative expansion the NJL model
becomes equivalent to the $\sigma$ model with lagrangian \be {\cal
L}_\sigma=\frac\beta 2\left( (\partial\sigma)^2+ (\partial{\bm
\pi})^2+\frac{\kappa^2}4\left(\sigma^2+{\bm\pi}^2-f_\pi^2\right)^2
\right)\ . \label{11}\ee

 The parameter $\beta$ is given by
\cite{Eguchi:1976iz}, \cite{Kleinert:1999wm}: \be
\beta=4g_0^2N_cN_f(I_0-2\Omega_0)\label{13}\ ,\ee with \bea
I_0&=&\int\frac{d^4p_E}{(2\pi)^4}\frac 1{(p_E^2+m^2_q)^2}\
,\cr&\cr \Omega_0&=&\frac 1 4 \int\frac{d^4p_E}{(2\pi)^4}\frac
{p_E^2}{(p_E^2+m^2_q)^3}\ .\label{14}\eea Classically one has
 $<\pi_a>=0$, $<\sigma>=f_\pi$; as a consequence the field
 $\sigma^\prime=\sigma\,-\,f_\pi$ acquires a mass $\kappa f_\pi/\sqrt 2$
 while the pions remain massless.

 If the $\sigma^\prime$ mass is
 very large, the model is equivalent to the non linear $\sigma$ model
 with action\be S=\frac{\beta}2\int d^4x \left((\partial\sigma)^2+ (\partial{\bm
\pi})^2\right)\ee and fields satisfying the constraint\be
\sigma^2+{\bm\pi}^2\,=\,f_\pi^2\ .\label{cond}\ee
\section{Stiffness\label{sec:3}}
 The generating functional of the nonlinear $\sigma$ model for vanishing external sources can be written
as follows\be Z=\int [d\sigma][ d{\bm\pi}][d\lambda]\exp\{iS\}\ee
\be S=\frac{\beta}2\int
d^Dx\left((\partial\sigma)^2+(\partial{\bm\pi})^2-\lambda
\left(\sigma^2+{\bm\pi}^2-f_\pi^2\right) \right)\ee where the
$\lambda$ functional integration implements the non linear
condition on the fields  (\ref{cond}). After  integration over the
${\bm\pi}$ fields we get \be Z=\int[d\sigma][d\lambda]\exp(
iS_{eff}[\sigma,\lambda])\ee with\be S_{eff}[\sigma,\lambda]=\frac
i 2 {\rm Tr}\ln[\partial^2+\lambda(x)]+\frac{\beta} 2\int d^D x
\left(f_\pi^2\lambda(x)-\sigma[\partial^2+\lambda(x)]\sigma\right)\
.\ee

 If the number of flavors is large we can use the saddle point
 approximation and search for solutions $\sigma,\,\lambda$ independent of
 $x$. Since we are looking for phase transitions at finite $T$ and $\mu$,
  we consider Matsubara frequencies and non vanishing baryonic chemical
  potential.
 One gets in this way \cite{Ripka:2000aa}:\bea
 0&=&\beta(f_\pi^2-\sigma^2)-(N_f^2-1)
\,\sum_{n=-\infty}^{\infty} \int\frac{d^3  k}{(2\pi)^3} \frac
T{(\tilde\omega_n-i\mu)^2+k^2+\lambda}\cr 0&=&\lambda\sigma\eea
with $\tilde\omega_n=2\pi nT$; therefore one has two solutions
 \cite{Ripka:2000aa}:\begin{itemize}
    \item i) $\sigma=0$ and $ \lambda$ implicitly given by \be\label{l}
    \beta f_\pi^2=(N_f^2-1)
\,\sum_{n=-\infty}^{\infty} \int\frac{d^3  k}{(2\pi)^3} \frac
T{(\tilde\omega_n-i\mu)^2+k^2+\lambda}\label{beta00}\ee
    \item ii) $\lambda=0$ and \be\sigma^2=f_\pi^2-\frac{N_f^2-1}{\beta}
\,\sum_{n=-\infty}^{\infty} \int\frac{d^3  k}{(2\pi)^3} \frac
T{(\tilde\omega_n-i\mu)^2+k^2}\ee
 \end{itemize}
 The case i) corresponds to chiral symmetry restoration, the case ii) to spontaneous symmetry
 breaking and massless pions. Note that in the chiral symmetric
 phase,  $\sigma=0$ and also $\langle 0| \bar q
q|0\rangle\,=\,0$.

Let us now consider the case of finite $T$ and $\mu$.
 The critical line of chiral symmetry restoration is defined by
 $\lambda=\sigma=0$, or \bea \beta_c& =&\frac{N_f^2-1}{f_\pi^2(T,\mu)}
\,\sum_{n=-\infty}^{\infty}\int\frac{d^3
 k}{(2\pi)^3}\frac T{(\tilde\omega_n-i\mu)^2+k^2}\cr&=&
\frac{N_f^2-1}{f_\pi^2(T,\mu)}
\frac{1}{8\pi^2}\int_0^{\Lambda_\pi}dk\,k\,\left[\coth\left(\frac{\mu+k}{2T}\right)-\coth\left(\frac{\mu-k}{2T}\right)\right]
\label{betac0} \eea
  At finite temperature and density we have for the stiffness,
  from Eqns. (\ref{13}), (\ref{14})
  \be
\beta=2g_0^2N_fN_c
T\sum_{n=-\infty}^{+\infty}\int_\Lambda\frac{d^3p}{(2\pi)^3}\left[\frac
1{[\epsilon^2+(\omega_n-i\mu)^2]^2}+\frac{m^2_q(T,\mu)}{[\epsilon^2+
(\omega_n-i\mu)^2]^3}\right]\label{betac}\ .\ee

 The critical temperature $T_c$ as a function of
$\mu$ is obtained by equating (\ref{betac0}) and(\ref{betac})
\cite{Kleinert:1999wm}, \cite{Babaev:2000fj}. The value of $T_c$
depends on the value of the pionic cutoff $\Lambda_\pi$; typical
values should be of a few units of  $ f_\pi$ since $f_\pi$ fixes
the scale of the derivative expansion. Numerically we find, at
$\mu=0$, for  $\Lambda_\pi=200 $ MeV, $T_c=146 $  MeV;  for
$\Lambda_\pi=300 $ MeV, $T_c=127$ MeV. We stress the assumption of
two different cutoffs, $\Lambda$ and $\Lambda_\pi$; for a
discussion on this point see \ref{conclusions}.

If $T_c(\mu)$ defines the critical line, we can compute  the mass
of the composite boson formed by the $q\bar q$ pair above
$T_c(\mu)$, i.e. in the Wigner phase. The existence of such pairs
in the NJL model is a feature that was observed long ago
\cite{Hatsuda:1985eb} and interpreted as precursor of chiral phase
transition at finite temperature and chemical potential. We differ
from these authors for two important aspects. First, we identify
the critical temperature as $T_c$ and not as $T^*$. Second, in
\cite{Hatsuda:1985eb} the fluctuations are identified by looking
at peaks in  the $\omega$ distribution of the strength function
\be {\cal A}({\bm k},\,\omega)\propto {\rm Im} G^R({\bm
k},\,\omega)\ee at ${\bm k}=0$ ($ G^R$ is the retarded Green
function describing the fluctuation of the order parameter in the
Wigner phase). Here we use the stiffness to take into account the
fluctuations and compute the mass of the $\pi$ mode
$m_\pi^2=\lambda$ by identifying $\beta(\lambda)$ in
(\ref{beta00}) with $\beta$ given by (\ref{betac}).

\begin{figure}
\begin{center}
\epsfxsize=9truecm \centerline{\epsffile{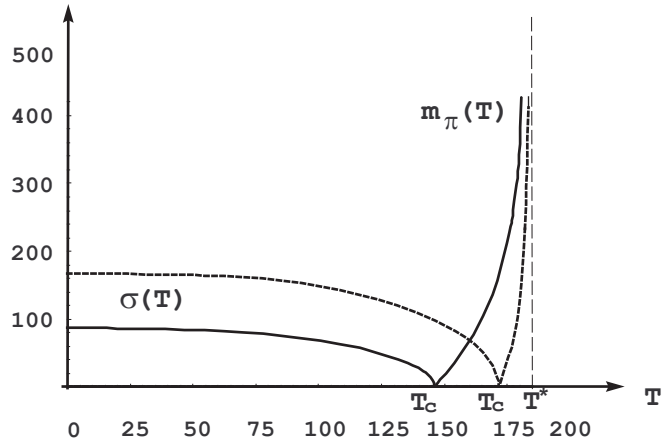}}
\noindent{\caption{ \label{fig:massa} -{\it{ The parameter $\sigma$
and mass of the pionic mode $m_\pi$ as functions of the temperature
$T$ at $\mu=0$ (in MeV). $T_c$ is the critical temperature. For
$T>T_c$ $\sigma$ vanishes, while $m_\pi$ vanishes for $T<T_c$.
Values of the parameters are $\Lambda_\pi=200$ MeV, $\Lambda=675$
MeV. Solid lines refer to the case $N_c=3$, dashed lines to $N_c=10$
(weaker coupling). As a result of the computation, $T_c\approx 146$
MeV for $N_c=3$ and $T_c\approx 172$ MeV for $N_c=10$. $T^*\simeq
T_{pair}$ is the dissociation temperature of the $\bar q q$
pairs.}}}}\end{center}
\end{figure}

For $\mu=0$ the results are reported in fig. \ref{fig:massa} for the
value of the parameter $\Lambda_\pi=200$ MeV and for $N_c=3$ (solid
lines) and $N_c = 10 $ (dashed lines). Before discussing the
dependence on $N_c$, let us note that in fig. \ref{fig:massa} the
function $\sigma(T)$ differs from zero in the interval $(0,T_c)$ and
vanishes in $(T_c,T^*)$ for both values of $N_c$. On the contrary
the function $m_\pi(T)$ vanishes in $(0,T_c)$ and rapidly increases
in $(T_c,T^*)$ with a divergent behavior around $T^*$.

An interesting aspect of these results is that they allow to define
a cross-over temperature $T_{pair}>T_c$; in our computation
$T_{pair}$ is the temperature at which the composite bosons
disappear from the physical spectrum.
As a result
of our approximation we find $T_{pair}\simeq T^*$. Lattice QCD
investigations show that the deconfinement and chiral phase
transitions take place at the same temperature, see e.g.
\cite{Karsch} and \cite{Bernard:1996iz}. Though our results indicate
the existence of two different temperatures, they do not contradict
lattice studies. As a matter of fact, the NJL model does not
incorporate confinement. Therefore we can only  state that our lower
critical temperature $T_c$ is the chiral restoration critical
temperature, while leaving undecided if it  coincides with the
deconfinement temperature, defined for example as the temperature
where the string tension vanishes. As for the upper characteristic
temperature $T_{pair} \simeq T^*$, it is unrelated with the
deconfinement temperature and is defined either as the temperature
where the strongly correlated Cooper pairs, behaving as massive
mesonic states in the interval $(T_c,T^*)$, decouple from the
spectrum, or as the temperature where the pseudogap phase
disappears.

We conclude that the behavior of this model is more akin to a
Bose-Einstein condensate than to a BCS system. Indeed while in the
latter case the temperature characterizing pair formation
coincides with the condensation temperature, the former case is
generally characterized by two critical temperatures, one (upper)
temperature when the pairs are preformed, and another (lower)
where the pairs condense (for discussions in condensed matter see
e.g. \cite{Eagles:1969} and for a review \cite{Loktev:2000ju}).
This behavior is confirmed by the large $N_c$ limit of the present
model. In this limit the four fermion coupling becomes weak, the
mean field approximation is exact and the two temperatures merge.
This behavior can be easily proved by taking the large $N_c$ limit
of the equation for the stiffness that determines $T_{pair}$. For
$N_c\to\infty$  one gets $T_c\simeq T_{pair}$, and the model
defines a BCS superconductor. By way of example we have reported
in Fig. \ref{fig:massa} the result of the calculations for
$N_c=10$ (dashed lines) showing that for weak coupling $T_c\to
T^*$.
\section{Pseudogap\label{sec:4}}The pseudogap can be
seen by the density distribution $N(\omega)$: \be
N(\omega)=4\int\frac{d{\bm k}}{(2\pi)^3} {\rm Tr}_{c,f}\rho^0({\bm
k},\omega) \ee ($c,f=$color, flavor), where\be\rho^0({\bm
k},\omega)=\frac 1 4  {\rm Tr}_{d} \gamma^0{\cal A}({\bm k},
\omega)\ee ($d=$Dirac) and \be {\cal A}({\bm k}, \omega)=-\frac 1
\pi {\rm Im} G^R({\bm k}, \omega)\ .\ee $G^R({\bm k}, \omega)$ is
the analytic continuation of the imaginary time Green function
${\cal G}$, in other words \be G^R({\bm k},
\omega)=\left[G^{-1}({\bm k},\omega+ i \epsilon)-\Sigma_R({\bm
k},\omega) \right]^{-1} \ee where $G^{-1}({\bm k},
\omega)=(\omega+\mu)\gamma^0-{\bm k}\cdot{\bm \gamma}-m_q$ is the
analytic continuation of the imaginary time Green function for
free quarks ${\cal G}^{-1}({\bm k},
\omega)=(i\omega_n+\mu)\gamma^0-{\bm k}\cdot{\bm \gamma}-m_q$.
Notice that including only the mass term, i.e. at the lowest order
in the interactions, we have\be
N(\omega)=N_{0}(\omega)=\frac{N_cN_f}{\pi^2}(\omega+\mu)\sqrt{(\omega+\mu)^2
-m_q^2}\ .\ee We include the quark mass term in $G({\bm k},
\omega)$ in the spirit of an effective lagrangian approach, where
the four-fermion coupling modifies the free quark dispersion law
and  the interactions are provided by the fluctuating pion field.
Actually the absence of interactions would imply also the
vanishing of $m_q$ and one would get, in this case, \be
N(\omega)=N_{free}(\omega)=
\frac{N_cN_f}{\pi^2}(\omega+\mu)^2\label{36}\ .\ee To compute the
effect of the interactions at the next order we consider the
effective lagrangian displayed in eq. (\ref{13bis}):\be{\cal
L}_{eff}=
\bar\psi[i\partial_\nu\gamma^\nu+\mu\gamma_0-g_0(\sigma+i\gamma_5{\bm
\tau}\cdot{\bm\pi})]\psi-\frac{g_0^2N_c}{2G_0}\left[\sigma^2+{\bm
\pi}^2\right]\ee together  with the condition (\ref{cond}). At the
lowest order in the pion field we have the following interaction
lagrangian \be {\cal L}_{int}=
\bar\psi\left(\frac{g_0}{2f_\pi}{\bm \pi}^2- ig_0\gamma_5{\bm
\tau}\cdot{\bm \pi}\right)\psi\ee together with the mass term
$-g_0f_\pi\bar\psi \psi= -m_q\bar\psi\psi$.
 At the lowest order in $g_0^2$ we  find
 \be \Sigma_R({\bm k},
\omega)=i\left((\alpha_1+i\alpha_2)\gamma_0+(\lambda_1+i\lambda_2)
{\bf k}\cdot{\bm\gamma}+\zeta_1+i\zeta_2\right)\ee and
($\alpha_i,\lambda_i\,,\zeta_i$ real): \bea
\alpha_1+i\alpha_2&=&-\,(N_f^2-1)\,
T\,\sum_{n=-\infty}^{+\infty}\,\int_{\Lambda_\pi}[dq]\,g_0^2({\bf
q}^2)\ \frac{i\omega_n-\omega-\mu}{D} \cr
\lambda_1+i\lambda_2&=&-\,(N_f^2-1)\,
T\,\sum_{n=-\infty}^{+\infty}\,\int_{\Lambda_\pi}[dq]\,g_0^2({\bf
q}^2)\ \frac{1-{\bf q}\cdot{\bf k}/{\bf k}^2}{D} \cr
\zeta_1+i\zeta_2&=&(N_f^2-1)\,
T\,\sum_{n=-\infty}^{+\infty}\,\int_{\Lambda_\pi}[dq]\,g_0^2({\bf
q}^2)\ \left[\frac{1}{2m_q}-\frac{m_q}{D}\right]\ .\label{35}\eea
The measure $[dq]$ is\be [dq]=\frac{d^3q}{(2\pi)^3}\frac{1}{(2\pi n
T)^2+{\bf q}^2+m_\pi^2} \ee and \be
D=[\omega_n+i(\omega+\mu)]^2+({\bf k}-{\bf q})^2+m^2_q\ee with
$\omega_n=2\pi T(n+1/2)$. We have included a form factor in the
definition of $g_0^2$:\be g_0^2({\bf
q}^2)=g_0^2\,\left(\frac{m_P^2}{m^2_P+(2\pi nT)^2+{\bf
q}^2}\right)^2\ .\ee It corrects the large $q^2$ behavior of the
loop diagrams and plays a role analogous to the regulator used in
\cite{Ripka:2000aa}. We take $m_P=100$ MeV, but the exact numerical
value is of no interest here as we are only interested in the
qualitative behavior of the density of states.
\begin{figure}[ht]
\begin{center}
\epsfxsize=9truecm
\centerline{\epsffile{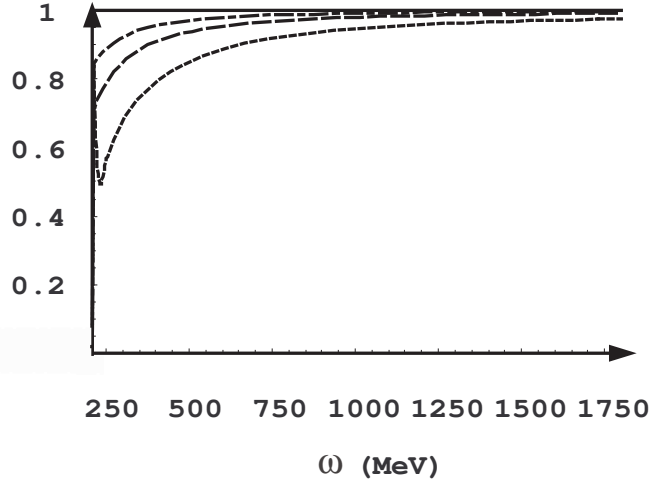}\hskip2cm}
\noindent{\caption{ \label{fig:ps} -{\it{ The ratio
$N(\omega)/N_{free}(\omega)$ as a function of the energy $\omega$.
$N(\omega)$ is the density of states for the interacting theory;
$N_{free}(\omega)$ the density for the free fields. Results are
presented for  $\mu=10$ MeV and for various temperatures. The
solid line represents the absence of interactions
 and corresponds to $T=T^*$. The other
curves refer to three different values of
$\,\dd\delta=\frac{T-T_c}{T_c}$: $\delta=0.02$, dotted;
$\delta=0.15$, dashed and  $\delta=0.20$, dotted-dashed line.
}}}}\end{center}
\end{figure}
We note that the effect of the form factor is to reduce the
perturbative contribution to the density distribution $N(\omega)$.
As a matter of fact this perturbative term decreases with decreasing
$m_P$, while for $m_P \to \infty$ the effect of the form factor
vanishes. Our numerical choice, $m_P=100 $ MeV, only indicates an
order of magnitude. Much larger values are excluded since they would
extend the validity of the approximation beyond the kinematical
range dictated by the neglect of higher order terms in the
derivative expansion of eq. (\ref{11}). We also note that a further
effect of the form factor is to make the series in eq. (\ref{35})
rapidly convergent, which allows an inversion between analytic
continuation and Matsubara summation, see e.g. ref. \cite{kitazawa},
and related work in refs. \cite{Kitazawa:20012003,Abuki:20012005}.

 Similarly to the
$N_{free}$ case, the $|{\bf k}|$ integration can be performed
using the residues theorem. Our results are as follows; we define
$N(\omega)=N_{0}(\omega)+N_{pert}(\omega)$ and we get\be
 N_{pert}(\omega)=\frac{N_c
N_f}{\pi^2}\left(k_0\alpha_2(\omega,k_0)+\frac{f(\omega,k_0)}{2k_0}+\frac{f^\prime(\omega,k_0)}{2}\right)\
.\ee Here  $f^\prime$ denotes partial derivation in the $k$
variable, \be k_0=\sqrt{(\omega+\mu)^2-m_q^2}\ee and \be
f(\omega,k)=-2(\omega+\mu)[(\omega+\mu)\alpha_2(\omega,k)+
k^2\lambda_2(\omega,k)+m_q\zeta_2(\omega,k)]\ .\ee Our results are
displayed in Fig. \ref{fig:ps} where we plot $\dd
\frac{N(\omega)}{N_{free}(\omega)}$  for three different values of
the ratio \be\delta=\frac{T-T_c}{T_c}\ ,\ee i.e. $\delta=0.02$
(dotted line) $\delta=0.15$ (dashed line) and $\delta=0.20$
(dotted-dashed line) and for one value of the chemical potential
($\mu=10$ MeV). For other values of $\mu$ the results are similar.
The figure
 shows the
existence of a pseudogap in the $(T_c,T^*)$ region. The pseudogap
is maximal near $T_c$ and decreases for $T_c\to T^*$, which can be
easily understood as the combined effect of the vanishing
constituent mass $m_q$ and the divergent behavior of $m_\pi$ for
$T\to T^*$.

\section{Conclusions and outlook\label{conclusions}}

The previous analysis shows that in the $T-\mu$ phase diagram of the
NJL model, at a fixed value of $\mu$ (smaller than the critical
value when the effects of color superconductivity become
significant), there are two typical temperatures, associated with
chiral symmetry restoration and with the dissociation of the $q \bar
q$ pairs. This qualitative behavior is suggested in
\cite{Babaev:2000fj} by  Babaev, who changes the conclusions of
\cite{Kleinert:1999wm} employing a similar method. The authors of
ref. \cite{Kleinert:1999wm} use $\Lambda_\pi\approx \Lambda$ and
conclude that the NJL model has no chiral broken phase, unless the
number of colors $N_c$ is larger than its physical value $N_c=3$. We
agree with  \cite{Babaev:2000fj} where two different ultraviolet
cutoffs are used, as they are related to different degrees of
freedom. Our explicit calculations corroborate therefore the
conclusions of \cite{Babaev:2000fj}. In particular we explicitly
prove the appearance of a pseudogap, related to the existence of a
strong coupling. The existence of two different cutoffs, one
fermionic ($\Lambda$) and another one bosonic ($\Lambda_\pi$) is due
to the fact that, as shown in refs. \cite{Dmitrasinovic:1995cb} and
\cite{Nikolov:1996jj}, the regularization of quark loops by means of
$\Lambda$ does not restrict the four momenta of the mesons. In all
non-renormalizable field theories, such as the NJL model, new
parameters are introduced as more loops are included in the
calculations. Since the inclusion of internal boson lines
corresponds to add more loops, the presence of meson propagators
implies the existence of the new cutoff $\Lambda_\pi$
\cite{Dmitrasinovic:1995cb}. In \cite{Nikolov:1996jj} the interval
(0, 1.5) for $\Lambda_\pi/\Lambda$ is investigated, but in
\cite{Oertel:1999fk} (see also \cite{Oertel2000}) it is shown that
$\Lambda_\pi>\Lambda$ leads to instabilities. Our choice
$\Lambda_\pi<\Lambda$ is consistent with \cite{Oertel:1999fk}: in
particular we find that the value $\Lambda_\pi=$200 MeV produces a
numerical value for the chiral phase transition in rough agreement
with lattice results. We finally note that the existence of two
different cutoffs can be also interpreted as the consequence of the
non-universality of a critical stiffness in the non linear sigma
model for 3+1 dimensions. This lack of universality implies
\cite{Babaev:2000fj} that the knowledge of $\beta_c$ in the NJL
model does not allow, by alone, to fix the position of the phase
transition in the effective field theory. Therefore the appearance
of a new parameter appears quite natural from this viewpoint.
Summarizing:  Our derivation is based, differently from Ref.
\cite{Kleinert:1999wm}, on two cutoffs and, as such, it might be
controversial. However, on the basis of the arguments of Refs.
\cite{Dmitrasinovic:1995cb,Nikolov:1996jj,Oertel:1999fk,Oertel2000}
and \cite{Babaev:2000fj} summarized above, we believe that using two
different cutoffs is a proper procedure.

 As already mentioned in
the introduction, some other models have been proposed to
understand in QCD the persistence of the $q \bar q$ bound states
above the critical temperature and the hydrodynamical properties
of the medium observed at RHIC; let us briefly comment on them. In
\cite{shu1} a strong QCD Coulomb-like interaction and the
formation of a multiple bound states of quasiparticles has been
suggested as a possible explanation of the data. However it is not
clear if this strong Coulomb-like interaction is obtained in
lattice simulations above the critical temperature \cite{karsh4}.
More recently in \cite{simonov} a strong non-confining  $q \bar q$
potential is predicted in the framework of the field correlator
method, which might account in part for the data.

The NJL model is obviously different from QCD; in particular it
does not contain one of the key dynamical QCD effects, i.e.
confinement. Nevertheless it contains the most relevant dynamics
of the effective four fermion interactions and is largely used as
an effective model of low energy QCD. Therefore the  similarity
between its behavior  and some non-perturbative features of QCD
(lattice results and the phenomenological interpretation of the
RHIC data) should not cause much surprise. On the basis of the
present work, it would be interesting to investigate if a
pseudogap also shows up in lattice QCD at intermediate
temperatures and small chemical potential. \vskip.2cm

{\bf Acknowledgements} We thank E. Babaev for useful
correspondence and  G. Angilella and V. Greco for useful
discussions.

\end{document}